\documentclass[10pt,aps,twocolumn,prl,noshowkeys,superscriptaddress,nofootinbib,longbibliography,floatfix]{revtex4-1}

\usepackage{amsmath,amsfonts,amsthm,amssymb,physics,graphicx}
\usepackage{newtxtext,newtxmath,booktabs,siunitx}
\usepackage[dvipsnames]{xcolor}
\usepackage[colorlinks=true,
			linktocpage=true,
			linkcolor=blue,
			citecolor=red,
			urlcolor=blue
            ]{hyperref}

\usepackage{bm}
\usepackage{enumitem}

\newcommand{\nn}{\nonumber\\ }
\newcommand{\nis}{\langle n_i\rangle_* }

\begin{document}

\title{Equilibrium Statistics as Conditional Laws and Conservation-Induced Correlations}

\author{Sunil Jaiswal}
\email{s.jaiswal@wayne.edu}
\affiliation{Department of Physics and Astronomy, Wayne State University, Detroit, Michigan 48201, USA}
\affiliation{Department of Physics, The Ohio State University, Columbus, Ohio 43210, USA}

\author{Amaresh Jaiswal}
\email{a.jaiswal@niser.ac.in}
\affiliation{School of Physical Sciences, National Institute of Science Education and Research, An OCC of Homi Bhabha National Institute, Jatni-752050, India}

\date{\today}


\begin{abstract} 
We present a novel unified conditional-probability framework for relativistic systems in which conditioning on additive conservation laws simultaneously yields equilibrium occupation statistics and conservation-induced correlations. In this formulation, equilibrium arises as a conditional limit law of a closed system. The one-mode marginal gives Maxwell--Boltzmann, Bose--Einstein, and Fermi--Dirac statistics at leading saddle order, with the conserved quantities fixing the exponential tilt and the microscopic occupation measure determining the statistics. Expanding the two-mode marginal to Gaussian order gives the leading finite-rank covariance between modes induced by exact conservation. When contracted with observables linear in mode occupations, this covariance gives their leading exact-conservation contribution. We use this structure to define projected observables orthogonal to selected conserved quantities. By construction, their covariance has no leading exact-conservation contribution. In small collision systems, where conservation effects are less suppressed by multiplicity and can survive standard nonflow suppressions, this provides a direct way to isolate conservation-aligned contributions to long-range correlations. We demonstrate this with PYTHIA8/Angantyr-generated p+Pb events at $\sqrt{s_{\mathrm{NN}}}=5.02~\mathrm{TeV}$ by comparing ordinary and projected covariances, showing that the projection removes the conservation-aligned contribution while leaving the conservation-orthogonal covariance essentially unchanged.
\end{abstract}


\maketitle


\textit{\textbf{Introduction ---}}
Long-range correlations observed in high-multiplicity p+p~\cite{CMS:2010ifv, ATLAS:2015hzw} and p+Pb collisions~\cite{ATLAS:2012cix, ALICE:2012eyl, CMS:2012qk} resemble correlation structures associated with collective dynamics in heavy-ion collisions~\cite{Heinz:2013th}, but appear in systems that are much smaller. This raises the question whether they reflect collective dynamics or arise from other long-range sources~\cite{Nagle:2018nvi, Grosse-Oetringhaus:2024bwr}. Interpreting these measurements requires separating collective dynamics from nonflow and finite-multiplicity effects. Experimentally, azimuthal and transverse-momentum correlations are often measured with pseudorapidity gaps or subevent methods to suppress short-range nonflow contributions~\cite{Jia:2017hbm, ATLAS:2017rtr, ALICE:2025iud, Wang:2026hak}. These procedures reduce local correlations from jets, resonance decays, and other short-range sources, but they do not remove correlations imposed by exact global conservation laws~\cite{Borghini:2000cm, Borghini:2002mv, Borghini:2003ur, Borghini:2006yk, Chajecki:2008vg}.  Conservation-induced correlations are intrinsically long range because particles in separated rapidity intervals must share the same globally conserved energy, momentum, and charges. Since conservation effects are less suppressed in small systems and peripheral events, they constitute an unavoidable contribution to measured long-range correlations~\cite{Vovchenko:2021yen, Poberezhnyuk:2022blc}. Existing treatments often compute exact-conservation effects as corrections to observables~\cite{Borghini:2002mv, Vovchenko:2021yen, Poberezhnyuk:2022blc}. Here we derive leading conservation-induced correlations for a general set of additively conserved quantities.

The derivation follows from formulating equilibrium statistics as conditional probability under additive conservation laws, rather than through ensembles, reservoirs, or entropy maximization~\cite{Kardar2007, pathriaSM2011}. This viewpoint is closely related to conditional limit theorems, the Gibbs conditioning principle, and large-deviation formulations of statistical mechanics~\cite{Campenhout1981, Csiszar1984, Diaconis1988, Ellis1985, Dembo1998, Ellis1999, Touchette2009, Touchette2015}. Here we formulate a many-mode system as an exact conditional probability law under additive conservation constraints. This provides a unified framework for equilibrium occupation statistics and conservation-induced correlations. The exact one-mode conditional law gives relativistic Maxwell--Boltzmann (MB), Bose--Einstein (BE), and Fermi--Dirac (FD) statistics at leading saddle order. The conserved quantities fix the common exponential tilt, while the allowed occupations and microscopic occupation weights determine the statistics. To best of our knowledge, this is the first derivation in which the relativistic Maxwell--J{\"u}ttner exponential form~\cite{Juttner1911, DeGroot:1980dk} arises from conditional probability, rather than by postulating a canonical ensemble or maximizing entropy.

The same conditional law also gives the correlations imposed by exact conservation. Expanding the exact two-mode conditional law to Gaussian order gives the leading conservation-induced covariance between modes. For observables linear in the mode occupations, this covariance gives the leading exact-conservation contribution to long-range correlations. It depends on each observable only through its overlap with the selected conserved quantities. This makes it possible to construct projected observables whose leading exact-conservation contribution vanishes by construction. We use this construction to define experimentally accessible long-range observables and demonstrate it with PYTHIA8/Angantyr-generated~\cite{Sjostrand:2014zea, Bierlich:2018xfw} p+Pb events at $\sqrt{s_{\mathrm{NN}}}=5.02~\mathrm{TeV}$. The projection removes the conservation-aligned part of the covariance while leaving the conservation-orthogonal covariance unchanged.

\textit{\textbf{Conditional framework ---}}
We consider a closed system with particles distributed among single-particle modes $i$. Let $n_i$ be the occupation number of mode $i$. We denote the
exactly conserved quantities by an $r$-component vector $Q$, with components
\begin{equation}
    Q^A=\sum_i q_i^A n_i \,,  \qquad A=1,\ldots,r \,,
\end{equation}
where $q_i^A$ is the $A$-th component of the conserved-quantity vector carried by mode $i$. A configuration $\{n_i\}$ is counted with microscopic weight $\mathbb{W}(\{n_i\})$, and the constrained density of configurations is
\begin{equation}
    \Omega(Q) = \sum_{\{n_i\}} \mathbb{W}(\{n_i\}) \delta^{(r)}\!\left(\sum_i q_i n_i-Q\right).  
    \label{eq:Omega}
\end{equation}
Here $\delta^{(r)}$ denotes the appropriate product of Dirac or Kronecker delta functions. We take the unconstrained reference measure to be independent across modes,
\begin{equation}
    \mathbb{W}(\{n_i\})=\prod_i W_i(n_i) \,.
    \label{eq:wf}
\end{equation}
Here $n_i\in \mathcal{A}_i$, where $\mathcal{A}_i$ is the set of allowed occupations of mode $i$. This factorization is appropriate for an ideal-gas reference measure, where the unconstrained occupation weights contain no intermode correlations. The allowed occupations $\mathcal{A}_i$ and the single-mode weights $W_i(n_i)$  distinguish the MB, BE, and FD cases.

\textit{\textbf{One-mode law and equilibrium statistics ---}}
Fixing the occupation $n_k$ of one mode forces the remaining modes to carry $Q-n_k q_k$. Therefore the exact one-mode conditional probability is
\begin{equation}
    \mathbb{P}_k(n_k\mid Q) = W_k(n_k) \frac{\Omega_{\neq k}(Q - n_k q_k)}{\Omega(Q)} \,.
    \label{eq:exact_one_mode}
\end{equation}
This exact identity is the starting point for the leading occupation laws. To evaluate the constrained densities, we introduce variables $\chi_A$ conjugate to the conserved quantities $Q^A$. The Laplace transform of $\Omega(Q)$ is
\begin{equation}
    \widehat{\Omega}(\chi) = \int d^r Q\, e^{-\chi_{_A} Q^A}\Omega(Q).
    \label{eq:laplace_def}
\end{equation}
For discrete conserved quantities, this notation denotes the corresponding generating function, and the inverse transform below should be understood as coefficient extraction. Using the definition of $\Omega(Q)$ and the factorized reference measure, Eqs.~\eqref{eq:Omega} and~\eqref{eq:wf}, one obtains
\begin{equation}
    \widehat{\Omega}(\chi) = \prod_i z_i(\chi), \quad
    z_i(\chi) = \sum_{n_i \in \mathcal{A}_i} W_i(n_i)e^{-n_i \chi_{_A} q_i^A} \,.
    \label{eq:sm_z_general}
\end{equation}
The constrained density is recovered from the inverse transform
\begin{equation}
    \Omega(Q) = \int_\Gamma \frac{d^r\chi}{(2\pi i)^r} \exp\!\left[\Phi(\chi;Q)\right],
    \label{eq:inverse_laplace}
\end{equation}
where the contour $\Gamma$ is chosen in the convergence domain of the transform, and $\Phi(\chi;Q) = \chi_{_A} Q^A + \sum_i \log z_i(\chi)$.
The constrained density is evaluated by expanding Eq.~\eqref{eq:inverse_laplace} around its stationary point $\chi_{_A}^*$. The stationarity condition fixes $\chi_{_A}^*$ through
\begin{equation}
    Q^A=\sum_i q_i^A \langle n_i\rangle_* \,, 
    \label{eq:saddle_constraints}
\end{equation}
where
\begin{equation}
    \langle n_i\rangle_* = \frac{1}{z_i(\chi^*)} \sum_{n_i \in \mathcal{A}_i} n_i W_i(n_i) e^{-n_i \chi_{_A}^* q_i^A} \,.
    \label{eq:nbar_saddle}
\end{equation}
Thus the saddle variables are fixed by the imposed conserved quantities.

At leading saddle order, the numerator in Eq.~\eqref{eq:exact_one_mode} is expanded around the saddle fixed by the total conserved quantities $Q$. The $n_k$-dependent part of the density-of-states ratio is
\begin{equation}
    \Omega_{\neq k}(Q - n_k q_k) \propto \frac{e^{-n_k \chi_{_A}^* q_k^A}}{z_k(\chi^*)} \,,
\end{equation}
where the proportionality constant is independent of $n_k$. Therefore
\begin{equation}
    \mathbb{P}_k^{(0)}(n_k\mid Q) = \frac{1}{z_k(\chi^*)} W_k(n_k) e^{-n_k \chi_{_A}^* q_k^A} \,.
    \label{eq:leading_one_mode}
\end{equation}
This is the conditional equilibrium law. The conserved quantities determine the exponential tilt through $\chi_{_A}^* q_k^A$, while the occupations $\mathcal{A}_k$ and weights $W_k(n_k)$ determine whether the resulting law is Maxwell--Boltzmann, Bose--Einstein, or Fermi--Dirac.

The stability of the saddle and the leading Gaussian response to conserved-quantity fluctuations are controlled by the Hessian of $\Phi$,
\begin{equation}
    H^{AB} = \left. \frac{\partial^2\Phi}{\partial \chi_{_A} \partial \chi_{_B}} \right|_{\chi^*}
    = \sum_i q_i^A q_i^B \sigma_{i,*}^2 \,.
    \label{eq:H}
\end{equation}
Here $\sigma_{i,*}^2 = \langle n_i^2 \rangle_* - \langle n_i\rangle_*^2$, evaluated with Eq.~\eqref{eq:leading_one_mode}. Thus $H^{AB}$ is the susceptibility matrix of the conserved quantities in the leading product measure. A regular saddle requires $H^{AB}$ to be nonsingular on the constrained subspace. Redundant conserved directions, or directions with zero fluctuations, should be removed before using the inverse susceptibility in the covariance projection below.

The usual equilibrium distributions follow directly from Eq.~\eqref{eq:leading_one_mode}. For four-momentum and particle-number constraints, $Q=(P^\mu,N)$ and $q_k=(p_k^\mu,1)$. Writing $\chi_{_A}^*=(\beta_\mu^*,-\alpha^*)$,  the saddle variable entering the one-mode law is
\begin{equation}
    x_k^* \equiv \chi_{_A}^* q_k^A = \beta_\mu^* p_k^\mu - \alpha^* \,.
\end{equation}
Here $\beta_\mu^*$ and $\alpha^*$ are the saddle values of the Laplace variables conjugate to $P^\mu$ and $N$. For an unbounded relativistic spectrum, convergence of the transform requires $\beta_\mu p^\mu>0$ for all future-directed momenta, so $\beta^\mu$ lies in the future-timelike domain. The leading conditional law becomes
\begin{equation}
    \mathbb{P}_k^{(0)}(n_k\mid P,N) = \frac{1}{z_k(x_k^*)} W_k(n_k) e^{-n_k x_k^*} \,.
    \label{eq:Pi0_xk}
\end{equation}
Above equation shows for the first time that the relativistic Maxwell–J{\"u}ttner exponential form~\cite{Juttner1911, DeGroot:1980dk} arises as a conditional tilt fixed by the imposed four-momentum and particle-number constraints, rather than by entropy maximization postulate.

For Maxwell--Boltzmann statistics~\cite{Maxwell1860,Boltzmann1877}, any number of particles may occupy the same mode, $\mathcal{A}_k=\{0,1,2,\ldots\}$, and the Gibbs factor removes overcounting by permutations, $W_k(n_k)=1/n_k!$~\cite{Gibbs1902}. Equation~\eqref{eq:Pi0_xk} then gives
\begin{equation}
    \langle n_k\rangle^{\rm MB}_* = \sum_{n_k=0}^{\infty} n_k\,\mathbb{P}_{k,{\rm MB}}^{(0)}(n_k \mid P,N) = e^{-x_k^*} \,.
    \label{eq:MB_result}
\end{equation}
%

For Bose--Einstein statistics~\cite{Bose1924,Einstein1924}, any number of identical bosons may occupy the same mode, $\mathcal{A}_k=\{0,1,2,\ldots\}$, and each occupation number has unit weight, $W_k(n_k)=1$. Equation~\eqref{eq:Pi0_xk} gives
\begin{equation}
    \langle n_k\rangle^{\rm BE}_* = \sum_{n_k=0}^{\infty} n_k\,\mathbb{P}_{k,{\rm BE}}^{(0)}(n_k \mid P,N) 
    = \frac{1}{e^{x_k^*}-1} \,.
    \label{eq:BE_result}
\end{equation}
%

For Fermi--Dirac statistics~\cite{Fermi1926,Dirac1926}, the Pauli exclusion rule restricts each mode to occupations $\mathcal{A}_k=\{0,1\}$, with unit weight $W_k(n_k)=1$. Equation~\eqref{eq:Pi0_xk} gives
\begin{equation}
    \langle n_k\rangle^{\rm FD}_* = \sum_{n_k=0}^{1} n_k\,\mathbb{P}_{k,{\rm FD}}^{(0)}(n_k \mid P,N)  
    = \frac{1}{e^{x_k^*}+1} \,.
    \label{eq:FD_result}
\end{equation}

For bosons, convergence requires  $x_k^*>0$ for each mode. In the local rest frame, $\beta^{*\mu}=u^\mu/T$ and $\alpha^*=\mu/T$, so $x_k^*=(E_k-\mu)/T$. The boundary $x_0\to0^+$ for the lowest mode corresponds to Bose accumulation~\cite{Bose1924, Einstein1924, Einstein1925}. In contrast, the FD occupation remains bounded and approaches the saturated Fermi surface as $T\to 0$~\cite{Sommerfeld1928}. Thus accumulation and saturation appear here as boundary behavior of the same conditional saddle.

\textit{\textbf{Two-mode law and conservation-induced correlations ---}}
We next apply the same conditional construction to two modes. For two distinct modes $k \neq \ell$, fixing $n_k$ and $n_\ell$ forces the remaining modes to carry $Q - n_k q_k - n_\ell q_\ell$. The exact joint conditional probability is therefore
\begin{equation}
    \mathbb{P}_{k\ell}(n_k, n_\ell \mid Q) = W_k(n_k) W_\ell(n_\ell) \frac{\Omega_{\neq k,\ell} (Q-n_k q_k - n_\ell q_\ell)} {\Omega(Q)} \,.
    \label{eq:exact_two_mode}
\end{equation}
This is the two-mode analogue of Eq.~\eqref{eq:exact_one_mode}. The leading one-mode law factorizes, but the exact constraint couples the two modes because their conserved-quantity fluctuations must be compensated by the remaining system.

A saddle expansion of the density-of-states ratio in Eq.~\eqref{eq:exact_two_mode} gives a Gaussian cost for the conserved quantity carried by the deviations $\Delta_i=n_i-\langle n_i\rangle_*$. The mixed part of this cost is
\begin{equation}
    -\Delta_k \Delta_\ell\, q_k^A (H^{-1})_{AB} \,q_\ell^B \,,
\end{equation}
where $H^{AB}$ is the susceptibility matrix in Eq.~\eqref{eq:H}. Since the leading law is factorized, this mixed term gives the leading connected covariance for $k \neq \ell$,
\begin{equation}
    \mathrm{Cov}_Q(n_k,n_\ell) = -\sigma_{k,*}^2 \sigma_{\ell,*}^2 q_k^A (H^{-1})_{AB}\, q_\ell^B +\cdots .
    \label{eq:offdiag_cov_main}
\end{equation}
The omitted terms include local one-mode corrections and higher-order terms in the saddle expansion. They do not change the leading off-diagonal covariance.

Combining this off-diagonal result with leading local variances gives the conditional-Gaussian projection form
\begin{equation}
    C_{ij}^{Q} = \sigma_{i,*}^2 \delta_{ij} - \sigma_{i,*}^2 \sigma_{j,*}^2 q_i^A (H^{-1})_{AB} \,q_j^B \,.
    \label{eq:C_projection}
\end{equation}
This form makes exact conservation manifest as $\sum_i q_i^A C_{ij}^{Q}=0$. Thus fluctuations along the exactly conserved directions are projected out. Equivalently, the conditional covariance has no component along the conserved directions. The detailed derivation is given in the Supplemental Material.

The same covariance form can be contracted with observables that are linear in the mode occupations. Let
\begin{equation}
    X=\sum_i f_i n_i,\qquad Y=\sum_i g_i n_i \,,
\end{equation}
where $f_i$ and $g_i$ are fixed coefficients that specify the selected modes and their relative weights in each observable. Contracting Eq.~\eqref{eq:C_projection} with these weights gives the conditional-Gaussian covariance
\begin{equation}
    \mathrm{Cov}^{\rm cg}_Q(X,Y) = \sum_i f_i g_i \sigma_{i,*}^2 - U_X^A (H^{-1})_{AB} U_Y^B ,
    \label{eq:linear_proj_main}
\end{equation}
where
\begin{equation}
    U_X^A=\sum_i f_i\sigma_{i,*}^2 q_i^A,\qquad
    U_Y^A=\sum_i g_i\sigma_{i,*}^2 q_i^A \,.
    \label{eq:U_def}
\end{equation}
The vectors $U_X^A$ and $U_Y^A$ measure the overlap of the observables with the conserved quantities. If $X$ and $Y$ are built from disjoint sets of modes, the local diagonal term in Eq.~\eqref{eq:linear_proj_main} is absent, and the conservation-induced covariance reduces to
\begin{equation}
    \mathrm{Cov}^{\rm cons}_Q(X,Y) = -U_X^A(H^{-1})_{AB}U_Y^B \,.
    \label{eq:linear_cons_main}
\end{equation}
Thus exact conservation produces a finite-rank covariance determined by the conserved-quantity overlaps $U_X^A$ and $U_Y^A$, and the inverse susceptibility matrix.

The finite-rank form of Eq.~\eqref{eq:linear_cons_main} gives a direct way to remove this leading conservation component. Suppose the projection is performed on the same set of modes used to define $H^{AB}$. For the observable $X=\sum_i f_i n_i$, we define projected coefficients
\begin{equation}
    f_i^\perp = f_i - q_i^A (H^{-1})_{AB} U_X^B \,.
    \label{eq:fperp}
\end{equation}
The corresponding projected observable is
\begin{equation}
    X^\perp = \sum_i f_i^\perp n_i \,.
    \label{eq:Xperp}
\end{equation}
By construction, $U_{X^\perp}^A=\sum_i f_i^\perp \sigma_{i,*}^2 q_i^A=0$.  Equation~\eqref{eq:linear_cons_main} gives $\mathrm{Cov}^{\rm cons}_Q(X^\perp,Y)=0$ for any disjoint linear observable $Y$, at this leading order and for the selected conserved quantities. Thus the projection removes only the weight component aligned with the selected conserved quantities. Consequently, the leading covariance generated by the selected global conservation laws vanishes, while local statistical and genuine dynamical correlations can remain.

\textit{\textbf{Conservation-orthogonal observables in high-energy collisions ---}}
We now specialize the projection to long-range correlation measurements in high-energy collisions. The aim is to construct modified analysis weights so that their overlap with a chosen set of exact conservation laws vanishes, and then compare the projected covariance with the ordinary covariance.

For an experimental analysis, a mode $i$ can be represented by an analysis bin. Such a bin may be specified by particle species, $p_T$, pseudorapidity, and azimuth, and $n_i$ denotes the event-by-event number of particles in that bin. Particles in bin $i$ are assigned conserved quantities $q_i^A$, such as momentum components $p_i^x, p_i^y$, energy $E_i$, electric charge, and other conserved charges. In a fixed multiplicity or centrality class $\mathcal{C}$, we write
\begin{equation}
    \bar{n}_i = \langle n_i\rangle_\mathcal{C},\qquad 
    \delta n_i=n_i - \bar{n}_i \,,
\end{equation}
where $\langle\cdots\rangle_\mathcal{C}$ denotes an event average over that class.

Let $L$ and $R$ be two non-overlapping rapidity-separated measured subevents in $\mathcal{C}$. The mean multiplicity in left subevent is $\overline{N}_L=\sum_{i\in L} \bar{n}_i$. We define the left-subevent fluctuation,
\begin{equation}
    X_L[f]=\frac{1}{\overline{N}_L}\sum_{i\in L} f_i\,\delta n_i \,.
    \label{eq:XL_HI}
\end{equation}
The coefficients $f_i$ define the observable. For example, $f_i=p_{T,i} - \bar{p}_{T,L}$, with $\bar{p}_{T,L} = \sum_{i\in L}p_{T,i} \bar{n}_i / \sum_{i\in L} \bar{n}_i$, gives a mean-$p_T$-type fluctuation, while $f_i=\cos(n\phi_i)$ or $\sin(n\phi_i)$ gives a harmonic weight. The overlap of this left-subevent observable with the conserved quantities is
\begin{equation}
    U_L^A=\sum_{i\in L} f_i d_i q_i^A \,.
    \label{eq:UL_HI}
\end{equation}
Here $d_i$ is the reference variance assigned to bin $i$. It plays the role of $\sigma_{i,*}^2$ in the formal projection~\eqref{eq:U_def} and determines how strongly bin $i$ enters the conserved-quantity overlap. The independent-particle choice $d_i=\bar n_i$ corresponds to a Poisson baseline for the bin occupation. In an experimental analysis, one may instead use measured or efficiency-corrected bin variances. The dependence on this choice can be used to estimate a systematic uncertainty of the projection. The right-subevent observable $X_R[g]$ and overlap $U_R^A$ are defined analogously. 

Since the measured particles are only a subset of the full event, the leading covariance induced by exact conservation is governed by the full-event susceptibility
\begin{equation}
    H_{\rm full}^{AB}=\sum_{a\in{\rm full\ event}} d_a\, q_a^A q_a^B \,,
    \label{eq:Hfull_HI}
\end{equation}
which is generally inaccessible from finite-rapidity data. The leading global-conservation contribution between the two measured subevents has the form~\eqref{eq:linear_cons_main}
\begin{equation}
    C_{\rm cons}^{LR}[f,g]=-\frac{1}{\overline{N}_L \overline{N}_R}\,U_L^A (H_{\rm full}^{-1})_{AB} U_R^B \,.
    \label{eq:LR_cons_HI}
\end{equation}
Thus the unknown full-event susceptibility is required to predict the magnitude of the conservation-induced covariance, but not to construct a projected observable for which the leading conservation contribution vanishes. If the measured overlaps $U_L^A$ and $U_R^A$ vanish for the conserved quantities being projected out, the leading global-conservation contribution vanishes independently of the unknown $H_{\rm full}^{AB}$.

We use only the measured subevent to define the weights
\begin{equation}
    S_L^{AB} = \sum_{i\in L} d_i q_i^A q_i^B \,.
    \label{eq:SL_HI}
\end{equation}
This matrix is not a full-event susceptibility and is used only to remove from $f_i$ the component aligned with the selected conserved quantities in the measured left-subevent. If some selected conserved direction is redundant in this measured set, the inverse is taken after removing that direction. The projected left-subevent coefficients are
\begin{equation}
    f_i^\perp = f_i - q_i^A (S_L^{-1})_{AB} U_L^B \,,
    \qquad i\in L  \,.
    \label{eq:fperp_HI}
\end{equation}
With this choice,
\begin{equation}
    U_{L,\perp}^A \equiv \sum_{i\in L} f_i^\perp d_i q_i^A=0 \,.
    \label{eq:Uperp_HI}
\end{equation}
Thus the projected observable has no overlap with the selected conserved quantities. Similarly, we define the right-subevent coefficients $g_j^\perp$. The projected subevent fluctuations are
\begin{equation}
    X_L^\perp=\frac{1}{\overline{N}_L}\sum_{i\in L}f_i^\perp\,\delta n_i \,, \qquad 
    X_R^\perp=\frac{1}{\overline{N}_R}\sum_{j\in R}g_j^\perp\,\delta n_j \,.
    \label{eq:Xperp_HI}
\end{equation}
The projected long-range covariance is
{\small
\begin{equation}
    C_\perp^{LR}[f,g] =\! \left\langle X_L^\perp[f]\,X_R^\perp[g] \right\rangle_{\mathcal C}
    \!\!= \frac{1}{\overline N_L\overline N_R} \sum_{i\in L}\sum_{j\in R} f_i^\perp g_j^\perp \!\left\langle \delta n_i\delta n_j\right\rangle_{\mathcal C}.
    \label{eq:Cperp_HI}
\end{equation}
}
Since $U_{L,\perp}^A = U_{R,\perp}^A = 0$, the leading exact-conservation contribution from the selected conserved quantities vanishes in $C_\perp^{LR}$. Note that the ordinary covariance,
\begin{equation}
    C^{LR}[f,g] = \mathrm{Cov}_\mathcal{C}\!\left(X_L[f], X_R[g]\right)
\end{equation}
contains the full measured long-range correlation, including any component aligned with the selected exact-conservation constraints. If exact conservation gives a significant contribution, the difference $C^{LR} - C_\perp^{LR}$ should increase toward lower multiplicity, where the same conserved quantities are shared among fewer particles. A signal surviving the projection is free of leading exact-conservation backgrounds from the selected conserved quantities, while retaining dynamical correlations such as jets, resonance decays, and local charge conservation.

\begin{figure}[t]
    \centering
    \includegraphics[width=\linewidth]{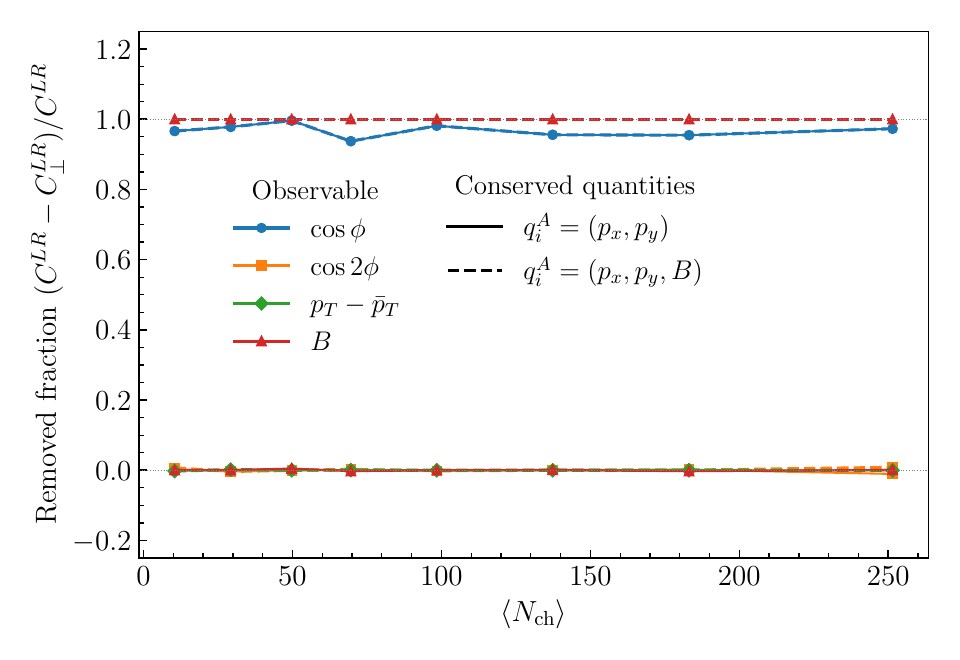}
    \vspace{-8mm}
    \caption{Projection test in PYTHIA8/Angantyr p+Pb at $\sqrt{s_{\mathrm{NN}}}=5.02~\mathrm{TeV}$. Charged final-state particles with $0.2<p_T<3~{\rm GeV}$ and $|\eta|<2.5$ are divided into two non-overlapping subevents, $-2.5<\eta<-0.8$ and $0.8<\eta<2.5$. The figure shows the removed fraction $(C^{LR}-C_\perp^{LR})/C^{LR}$ as a function of charged-particle multiplicity. Colors denote observable choice $f_i$, and line styles denote the conserved quantities included in the projection.}
    \label{fig:projection}
    \vspace{-4mm}
\end{figure}

We show the implementation in Fig.~\ref{fig:projection}, using $10^5$ PYTHIA8/Angantyr~\cite{Sjostrand:2014zea, Bierlich:2018xfw} p+Pb events. This analysis provides a closure test in a controlled event sample where the projection can be checked against transverse-momentum recoil and baryon-number conservation. The projection is applied separately in two rapidity-separated subevents, with $d_i=\bar{n}_i$ and bins in $p_T$ and $\phi$. Projection against $q_i^A=(p_{x,i},p_{y,i})$ removes the first-harmonic covariance, as expected from transverse-momentum recoil, while leaving the second harmonic and mean-$p_T$ covariance unchanged. For the baryon-number observable $f_i=B_i$, projecting only $(p_x,p_y)$ leaves the covariance unchanged, while including $B_i$ among the projected quantities removes it. Thus projection removes correlations aligned with selected conservation laws, while leaving conservation-orthogonal long-range components unchanged.

\textit{\textbf{Discussion —}}
The conditional framework presented here separates two aspects in equilibrium statistics. Conditional-limit reasoning explains why additive constraints produce an exponential tilt~\cite{Ellis1985, Touchette2009}. The microscopic occupation measure then determines the statistics. Poisson weights give MB statistics, unrestricted unit weights give BE statistics, and the exclusion constraint gives FD statistics. The Gaussian expansion around the same conditional saddle gives the susceptibility matrix that controls finite-size fluctuations and conservation-induced correlations.

The same framework turns exact conservation laws into projection variables. In high-energy collisions, the projection separates the component of a long-range correlation aligned with selected exact conservation laws from the conservation-orthogonal component. This is useful in small systems and peripheral collisions, where conservation effects are less suppressed by multiplicity. The construction can also be applied to conserved-charge fluctuation measurements near the QCD critical point, where global charge conservation must be separated from fluctuations generated by critical dynamics~\cite{Stephanov:1998dy, Poberezhnyuk:2020ayn}. Such a separation is crucial for the reliable identification of critical-point signatures. Few-body ultracold-atom expansions provide another setting where collective-like behavior has been reported in systems with very small particle number~\cite{Floerchinger:2021ygn, Brandstetter:2023jsy}. In such systems, correlations implied by fixed particle number, conserved energy--momentum, and fixed internal-state populations can be comparable to dynamical correlations. Applying the projection would test which part of the observed collective-like response survives after the leading exact-conservation component is removed. The framework developed here opens new avenues for identifying genuine collective dynamics in finite many-body systems.

\textit{\textbf{Acknowledgments ---}}
We thank Richard Furnstahl, Chun Shen, Soumangsu Chakraborty, Trakshu Sharma, and Chandrodoy Chattopadhyay for engaging discussions. S.J. gratefully acknowledges support from the CSSI program under Award No.~OAC-2004601 (BAND Collaboration~\cite{BAND_Framework}). S.J. acknowledges the warm hospitality of NISER Bhubaneswar where this work was initiated. A.J. acknowledges Department of Atomic Energy (DAE), India for financial support. 

\bibliography{ref}


\onecolumngrid
\newpage

\begin{center}
\underline{\bfseries\Large Supplemental Material}\\[0.5cm]

{\bfseries\large
Equilibrium Statistics as Conditional Laws and Conservation-Induced Correlations
}\\[0.3cm]

Sunil Jaiswal$^{1,2}$ and Amaresh Jaiswal$^{3}$\\[0.2cm]

{\itshape
$^{1}$Department of Physics and Astronomy, Wayne State University, Detroit, Michigan 48201, USA\\
$^{2}$Department of Physics, The Ohio State University, Columbus, Ohio 43210, USA\\
$^{3}$School of Physical Sciences, National Institute of Science Education and Research, An OCC of Homi Bhabha National Institute, Jatni-752050, India
}
\end{center}


\setcounter{equation}{0}
\setcounter{figure}{0}
\setcounter{table}{0}
\setcounter{page}{1}
\renewcommand{\theequation}{S\arabic{equation}}
\renewcommand{\thefigure}{S\arabic{figure}}

\maketitle

\vspace{1.0cm}


We present the derivation of the equations appearing in the main text. 

\section{Notation}

We denote the exactly conserved quantities by an $r$-component vector $Q$, with components
\begin{equation}
    Q^A=\sum_i q_i^A n_i, \qquad A=1,\ldots,r .
\end{equation}
Here $n_i$ is the occupation number in mode $i$, and $q_i^A$ is the $A$th component of the corresponding single-particle conserved-charge vector $q_i$. Capital indices $A,B,\ldots$ label the space of conserved quantities and are summed over when repeated: $\chi_{_A} q_i^A = \sum_{A=1}^r \chi_{_A} q_i^A$. Mode labels $i,j,\ldots$ are summed or multiplied over only when an explicit $\sum_i$ or $\prod_i$ is written. They should not be confused with Lorentz indices, which are denoted by Greek letters $\mu,\nu,\ldots$.

For the four-momentum and particle-number case used in the main text, we may take
\begin{equation}
    Q=(P^\mu, N), \qquad q_i=(p_i^\mu, 1).
\end{equation}
The last component of $q_i$ is unity because each particle occupying mode $i$ contributes one unit to the total particle number $N$. The corresponding conjugate variables are $\chi_{_A}=(\beta_\mu, -\alpha)$, so that
\begin{equation}
    \chi_{_A} Q^A = \beta_\mu P^\mu - \alpha N,
    \qquad
    \chi_{_A} q_i^A = \beta_\mu p_i^\mu - \alpha .
\end{equation}
For a hadron gas with exact four-momentum, baryon number, strangeness, and other conserved charges, one may instead take $Q=(P^\mu, B, S, \ldots)$ with $q_i=(p_i^\mu, B_i, S_i, \ldots)$. The signs of the corresponding chemical potentials are absorbed into the definition of $\chi_{_A}$.

\section{Equilibrium statistics as conditional limit laws}

We consider occupation configurations $\{n_i\}$, with allowed occupations $n_i \in \mathcal{A}_i$. A configuration $\{n_i\}$ is counted with a weight $W(\{n_i\})$, which encodes the microscopic counting. The constrained density of configurations at fixed $Q$ is
\begin{equation}
    \Omega(Q) = \sum_{\{n_i\}} \mathbb{W}(\{n_i\})\, \delta^{(r)} \left( \sum_i q_i n_i-Q \right).
    \label{eqA:Omega_Q}
\end{equation}
The notation $\delta^{(r)}$ denotes the appropriate product of Dirac or Kronecker delta functions, depending on whether the corresponding conserved quantities are continuous or discrete. For example, if $Q=(P^\mu,N)$ and $q_i=(p_i^\mu,1)$, then
\begin{equation*}
    \delta^{(r)}\left(\sum_i q_i n_i-Q\right) = \delta^{(4)}\left(\sum_i p_i^\mu n_i-P^\mu\right) \delta_{\sum_i n_i,N}.
\end{equation*}
For the Maxwell--Boltzmann (MB), Bose--Einstein (BE), and Fermi--Dirac (FD) counting rules considered in this work, the unconstrained microscopic counting assigns a separate weight to each mode. Hence
\begin{equation}
    \mathbb{W}(\{n_i\}) = \prod_i W_i(n_i).
    \label{eqA:weight_factorized}
\end{equation}

\subsection{Exact single-mode conditional distribution}

The factorization in Eq.~\eqref{eqA:weight_factorized} allows us to isolate the exact conditional distribution of a single mode. If mode $k$ has occupation $n_k \in \mathcal{A}_k$, the remaining modes must carry the conserved quantity $Q - n_k q_k$. Therefore the exact conditional probability for the occupation of mode $k$ is
\begin{equation}
    \mathbb{P}_k(n_k \mid Q) = W_k(n_k) \frac{\Omega_{\neq k}(Q-n_k q_k)}{\Omega(Q)} .
\label{eqA:exact_marginal_Q}
\end{equation}
Here $\Omega_{\neq k}$ denotes the constrained density of configurations of all modes except $k$. For the four-momentum and particle-number case, $Q=(P^\mu,N)$ and $q_k=(p_k^\mu,1)$, Eq.~\eqref{eqA:exact_marginal_Q} becomes
\begin{equation}
    \mathbb{P}_k(n_k\mid P,N) = W_k(n_k) \frac{\Omega_{\neq k}(P-n_k p_k,N-n_k)}{\Omega(P,N)} .
\end{equation}

\subsection{Laplace transform and conditional saddle}

To evaluate the constrained densities in the large-system limit, we introduce variables $\chi_{_A}$ conjugate to the conserved quantities $Q^A$. We define
\begin{equation}
    \widehat{\Omega}(\chi) = \int d^r Q\, e^{-\chi_{_A} Q^A}\Omega(Q).
    \label{eqA:Laplace_def}
\end{equation}
For conserved quantities with discrete components, this notation should be read as the corresponding generating function. The inverse transform then amounts to coefficient extraction in those discrete variables.

Using Eqs.~\eqref{eqA:Omega_Q} and \eqref{eqA:weight_factorized}, we obtain
\begin{align}
    \widehat\Omega(\chi) &= \sum_{\{n_i\}} \left[ \prod_i W_i(n_i)\right] \int d^r Q\, e^{-\chi_{_A} Q^A} \delta^{(r)} \left( \sum_i q_i n_i-Q \right)
    \nn
    &= \sum_{\{n_i\}} \left[ \prod_i W_i(n_i)\right] \exp \left( -\chi_{_A}\sum_i q_i^A n_i \right)
\end{align}
Since
\begin{equation}
    \exp \left( -\chi_{_A}\sum_i q_i^A n_i \right) =  \prod_i \exp\left(-n_i\chi_{_A} q_i^A\right),
\end{equation}
we have
\begin{align}
    \widehat\Omega(\chi)  = \sum_{\{n_i\}} \prod_i \left[W_i(n_i)\, \exp\left(-n_i\chi_{_A} q_i^A\right)\right]
    = \prod_i \sum_{n_i\in\mathcal A_i} W_i(n_i) \, \exp\left(-n_i\chi_{_A} q_i^A\right)
    = \prod_i z_i(\chi) \,.
    \label{eqA:Laplace_factorization}
\end{align}
Here we have defined the single-mode generating function
\begin{equation}
    z_i(\chi) = \sum_{n_i\in\mathcal A_i} W_i(n_i)\, e^{-n_i\chi_{_A} q_i^A} \,.
    \label{eqA:single_mode_z}
\end{equation}
The constrained density is recovered by the inverse transform
\begin{equation}
    \Omega(Q) = \int_{\Gamma} \frac{d^r\chi}{(2\pi i)^r} \exp\left[\Phi(\chi;Q)\right],
    \label{eqA:inverse_transform}
\end{equation}
with
\begin{equation}
    \Phi(\chi;Q) = \chi_{_A} Q^A + \sum_i \log z_i(\chi).
    \label{eqA:Phi_def}
\end{equation}
The contour $\Gamma$ is a product of Bromwich contours chosen within the convergence domain of the transform.

The constrained density is obtained by a saddle expansion of the inverse-transform integral. The dominant contribution comes from the stationary point $\chi_{_A}^*$ of the exponent $\Phi(\chi;Q)$, whose saddle condition is
\begin{equation}
    0= \left. \frac{\partial\Phi}{\partial\chi_{_A}} \right|_{\chi^*}
    = Q^A + \sum_i \left. \frac{\partial\log z_i}{\partial\chi_{_A}} \right|_{\chi^*}.
    \label{eqA:saddle_cond}
\end{equation}
Using Eq.~\eqref{eqA:single_mode_z}, 
\begin{equation}
    \frac{\partial\log z_i}{\partial\chi_{_A}} = -q_i^A
    \frac{\sum_{n_i\in\mathcal{A}_i} n_i W_i(n_i)e^{-n_i \chi_{_C} q_i^C}}{\sum_{n_i\in\mathcal{A}_i} W_i(n_i) e^{-n_i \chi_{_C} q_i^C}} \,.
    \label{eqA:dlogz}
\end{equation}
Thus the saddle equation becomes
\begin{equation}
    Q^A=\sum_i q_i^A \nis,
    \label{eqA:saddle_Q}
\end{equation}
where
\begin{equation}
    \nis = \frac{1}{z_i(\chi^*)} \sum_{n_i\in\mathcal{A}_i} n_i W_i(n_i)e^{-n_i \chi_{_A}^* q_i^A} \,.
\end{equation}

\subsection{Leading occupation law}

The leading one-mode law follows from the saddle expansion of the exact density-of-states ratio in Eq.~\eqref{eqA:exact_marginal_Q}. For fixed $n_k\in\mathcal{A}_k$, the numerator is evaluated at the shifted conserved quantity $Q-n_k q_k$. At leading saddle order, we evaluate the numerator and denominator at the same stationary point $\chi^*$. 

The numerator is then approximated by
\begin{equation}
    \Omega_{\neq k}(Q-n_k q_k) \simeq \exp\left[ \Phi_{\neq k}(\chi^*; Q-n_k q_k) \right],
\end{equation}
where
\begin{align}
    \Phi_{\neq k}(\chi^*; Q-n_k q_k) &= \chi_{_A}^* Q^A + \sum_{i\neq k}\log z_i(\chi^*) - n_k \chi_{_A}^* q_k^A 
    = \chi_{_A}^* Q^A + \sum_{i}\log z_i(\chi^*) - \log z_k(\chi^*) - n_k \chi_{_A}^* q_k^A 
    \nn
    &= \Phi(\chi^*;Q) - \log z_k(\chi^*) - n_k \chi_{_A}^* q_k^A .
\end{align}
Similarly, the denominator is approximated by
\begin{equation}
    \Omega(Q) \simeq \exp\left[\Phi(\chi^*;Q)\right].
\end{equation}
Substituting these into Eq.~\eqref{eqA:exact_marginal_Q}, we obtain
\begin{equation}
    \mathbb{P}_k^{(0)}(n_k \mid Q) =   W_k(n_k) \left. \frac{\Omega_{\neq k}(Q-n_k q_k)}{\Omega(Q)} \right|_{\chi^*}
    = \frac{1}{z_k(\chi^*)} W_k(n_k) e^{-n_k\chi_{_A}^* q_k^A}.
    \label{eqA:Pk0_def}
\end{equation}
The exponential factor is determined by the saddle values $\chi_{_A}^*$, which are fixed by the imposed conserved quantities through Eq.~\eqref{eqA:saddle_Q}. 
For the four-momentum and particle-number case, $Q=(P^\mu,N)$ and $q_k=(p_k^\mu,1)$, we write $\chi_{_A}^*=(\beta_\mu^*,-\alpha^*)$. Here $\beta_\mu$ and $\alpha$ are Laplace variables conjugate to $P^\mu$ and $N$. For an unbounded relativistic spectrum, convergence requires $\beta_\mu p^\mu>0$ for all future-directed on-shell momenta $p^\mu$, which places $\beta^\mu$ in the future-timelike domain. 

\clearpage
\subsection{Hessian and saddle stability}

The stability of the conditional saddle is determined by the quadratic variation of the exponent around the stationary point. We write
\begin{equation}
    \chi_{_A} = \chi_{_A}^* + \delta \chi_{_A} .
\end{equation}
Expanding the exponent about the stationary point gives
\begin{equation}
    \Phi(\chi;Q) = \Phi(\chi^*;Q) + \frac{1}{2}\, \delta\chi_{_A} H^{AB}\delta\chi_{_B} +\cdots ,
    \label{eqA:Phi_gaussian_general}
\end{equation}
where the linear term is absent by the saddle equation~\eqref{eqA:saddle_cond}, and
\begin{equation}
    H^{AB} \equiv \left. \frac{\partial^2\Phi}{\partial\chi_{_A} \partial\chi_{_B}} \right|_{\chi^*} \,,
    \label{eqA:Hessian_def_general}
\end{equation}
is the Hessian of the exponent at the saddle. Since $\Phi(\chi;Q)=\chi_{_A}Q^A+\sum_i\log z_i(\chi)$, 
\begin{align}
    H^{AB} &= \left.\frac{\partial}{\partial \chi_{_B}} \left(\frac{\partial \Phi}{\partial\chi_{_A}} \right) \right|_{\chi^*} 
    = \left.\frac{\partial}{\partial \chi_{_B}} \left(Q^A + \sum_i \frac{\partial \log z_i}{\partial\chi_{_A}} \right) \right|_{\chi^*} 
    = \left. \sum_i \frac{\partial^2 \log z_i}{\partial\chi_{_A}  \partial\chi_{_B}} \right|_{\chi^*}
\end{align}
Using Eq.~\eqref{eqA:single_mode_z}, the first derivative is
\begin{equation}
    \frac{\partial \log z_i}{\partial \chi_{_A}} = -q_i^A \langle n_i\rangle ,
\end{equation}
where
\begin{equation}
    \langle n_i\rangle = \frac{1}{z_i(\chi)} \sum_{n_i\in\mathcal A_i} n_i W_i(n_i)e^{-n_i\chi_{_C} q_i^C}.
\end{equation}
Taking one more derivative gives
\begin{align}
    \frac{\partial^2 \log z_i}{\partial \chi_{_A} \partial\chi_{_B}} &= -q_i^A \frac{\partial \langle n_i\rangle}{\partial\chi_{_B}}
    = -q_i^A \frac{\partial}{\partial\chi_{_B}} \left[ \frac{ \sum_{n_i\in\mathcal A_i} n_i W_i(n_i)e^{-n_i\chi_{_C} q_i^C} }{ z_i(\chi) } \right]
    \nn
    &= -q_i^A \left[ \frac{1}{z_i(\chi)} \frac{\partial}{\partial\chi_{_B}} \sum_{n_i\in\mathcal A_i} n_i W_i(n_i)e^{-n_i\chi_{_C} q_i^C} 
    - \frac{ \sum_{n_i\in\mathcal A_i} n_i W_i(n_i)e^{-n_i\chi_{_C} q_i^C} }{ z_i(\chi)^2 } \frac{\partial z_i(\chi)}{\partial\chi_{_B}} \right]
    \nn
    &= -q_i^A \left[ -q_i^B \langle n_i^2\rangle + q_i^B \langle n_i\rangle^2 \right]
    = q_i^A q_i^B \left( \langle n_i^2\rangle-\langle n_i\rangle^2 \right).
\end{align}
Therefore, at the saddle,
\begin{equation}
    H^{AB} = \left. \frac{\partial^2\Phi}{\partial\chi_{_A}\partial\chi_{_B}} \right|_{\chi^*} = \sum_i q_i^A q_i^B \sigma^2_{i,*},
    \label{eqA:Hessian_general}
\end{equation}
where
\begin{equation}
    \sigma^2_{i,*}  = \langle n_i^2\rangle_* - \langle n_i\rangle_*^2 
\end{equation}
is the variance of the leading one-mode law in Eq.~\eqref{eqA:Pk0_def}. 

Equivalently, we can define the fluctuation of the conserved quantities in the leading product measure by
\begin{equation}
    \delta Q^A = \sum_i q_i^A \left( n_i-\langle n_i\rangle_* \right),
\end{equation}
Then
\begin{equation}
    H^{AB} = \left\langle \delta Q^A \delta Q^B \right\rangle_* .
    \label{eqA:Hessian_as_covariance}
\end{equation}
Thus $H^{AB}$ is the covariance matrix of the conserved quantities before the exact constraint is imposed. The inverse of $H^{AB}$ exists only if the imposed conserved quantities are independent and have nonzero fluctuations. In other words, for every nonzero vector $v_A$ in the conserved-charge space,
\begin{equation}
    v_AH^{AB}v_B = \sum_i \sigma_{i,*}^2 \left( v_A q_i^A \right)^2 > 0 .
    \label{eqA:Hessian_positive_condition}
\end{equation}
This condition is the saddle-stability condition in the constrained directions. It fails if one of the constraints is redundant, or if some imposed conserved combination does not fluctuate in the leading product measure. In that case the corresponding direction must be removed before using the projection formula.

When Eq.~\eqref{eqA:Hessian_positive_condition} holds, we define the inverse Hessian as
\begin{equation}
    H^{AC}(H^{-1})_{CB} = \delta^A_{\ B}.
    \label{eqA:H_inverse_def}
\end{equation}
The position of capital indices is not tensorial. $H^{AB}$ and $(H^{-1})_{AB}$ denote a matrix and its inverse.

\subsection{Equilibrium statistics}

For the four-momentum and particle-number constraint, we take
\begin{equation}
    Q=(P^\mu,N), \qquad q_k=(p_k^\mu,1),
    \qquad \chi_{_A}^*=(\beta_\mu^*,-\alpha^*) .
\end{equation}
The leading one-mode law given in Eq.~\eqref{eqA:Pk0_def} becomes
\begin{equation}
    \mathbb{P}_k^{(0)}(n_k\mid P,N) = \frac{1}{z_k(\beta^*,\alpha^*)} W_k(n_k) \exp\left[-n_k(\beta_\mu^*p_k^\mu-\alpha^*)\right].
    \label{eqA:Pk0_PN}
\end{equation}
Defining
\begin{equation}
    x_k \equiv \beta_\mu^*p_k^\mu-\alpha^* ,
\end{equation}
the single-mode generating function is
\begin{equation}
    z_k(\beta^*,\alpha^*) = \sum_{n_k\in\mathcal A_k} W_k(n_k)e^{-n_k x_k}.
\end{equation}

For Maxwell--Boltzmann counting, $\mathcal A_k=\{0,1,2,\ldots\}$ and $W_k(n_k)=1/n_k!$. Hence
\begin{align}
    \langle n_k\rangle^{\rm MB}_* &= \sum_{n_k=0}^{\infty} n_k\,\mathbb P_{k,{\rm MB}}^{(0)}(n_k \mid P,N)  
    = \frac{\sum_{n_k=0}^{\infty} n_k\,e^{-n_k x_k}/n_k!}{\sum_{n_k=0}^{\infty} e^{-n_k x_k}/n_k!} 
    = e^{-x_k} \,.
\end{align}

For Bose--Einstein counting, $\mathcal A_k=\{0,1,2,\ldots\}$ and $W_k(n_k)=1$. Therefore
\begin{align}
    \langle n_k\rangle^{\rm BE}_* &= \sum_{n_k=0}^{\infty} n_k\,\mathbb P_{k,{\rm BE}}^{(0)}(n_k \mid P,N) 
    = \frac{\sum_{n_k=0}^{\infty} n_k\,e^{-n_k x_k}}{\sum_{n_k=0}^{\infty} e^{-n_k x_k}} 
    = \frac{1}{e^{x_k}-1} \,.
\end{align}
Note that the sums converge only for $x_k>0$.

For Fermi--Dirac counting, $\mathcal A_k=\{0,1\}$ and $W_k(n_k)=1$. Thus
\begin{align}
    \langle n_k\rangle^{\rm FD}_* &= \sum_{n_k=0}^{1} n_k\,\mathbb P_{k,{\rm FD}}^{(0)}(n_k \mid P,N) 
    = \frac{\sum_{n_k=0}^{1} n_k\,e^{-n_k x_k}}{\sum_{n_k=0}^{1} e^{-n_k x_k}}
    = \frac{1}{e^{x_k}+1} \,.
\end{align}

\clearpage
\section{Two-mode conditional law and induced covariance}

We now derive the leading correlation between two modes induced solely by the exact conservation constraint. Consider two distinct modes $k\neq \ell$, with allowed occupations $n_k \in \mathcal{A}_k$ and $n_\ell \in \mathcal{A}_\ell$. If these occupations are fixed, the remaining modes must carry the conserved quantity $Q-n_k q_k-n_\ell q_\ell$. Using the factorized microscopic weight, the exact joint conditional probability is
\begin{equation}
    \mathbb{P}_{k\ell}(n_k,n_\ell\mid Q) = W_k(n_k) W_\ell(n_\ell) \frac{\Omega_{\neq k,\ell}(Q-n_k q_k-n_\ell q_\ell)}{\Omega(Q) } .
    \label{eqA:two_mode_exact}
\end{equation}
Here $\Omega_{\neq k,\ell}$ denotes the constrained density of configurations of all modes except $k$ and $\ell$. 

\subsection{Saddle expansion of the two-mode law}

We now expand both the numerator and denominator in Eq.~\eqref{eqA:two_mode_exact} around the same full-system saddle $\chi^*$. The denominator is written as the inverse transform
\begin{equation}
    \Omega(Q) = \int_{\Gamma}\frac{d^r\chi}{(2\pi i)^r} \exp\left[\Phi(\chi;Q)\right],
    \label{eqA:TM_denominator_inverse}
\end{equation}
with
\begin{equation}
    \Phi(\chi;Q) = \chi_{_A}Q^A+\sum_i\log z_i(\chi).
    \label{eqA:TM_Phi_denominator}
\end{equation}
Introducing fluctuations around the saddle,
\begin{equation}
    \chi_{_A}=\chi_{_A}^* + \delta\chi_{_A} ,
\end{equation}
the denominator exponent expands as
\begin{equation}
    \Phi(\chi;Q) = \Phi(\chi^*;Q) + \frac{1}{2}\delta\chi_{_A} H^{AB}\delta\chi_{_B} +\cdots ,
    \label{eqA:TM_denominator_Taylor}
\end{equation}
where the linear term is absent by the saddle equation, and $H^{AB}$ is the Hessian defined in Eq.~\eqref{eqA:Hessian_general}. Near the saddle, the inverse-transform contour is locally parametrized by
\begin{equation}
    \delta\chi_{_A}=i y_A ,
\end{equation}
with real $y_A$. The denominator exponent then becomes
\begin{equation}
    \Phi(\chi;Q) = \Phi(\chi^*;Q) - \frac{1}{2}y_AH^{AB}y_B+\cdots .
    \label{eqA:TM_denominator_gaussian_y}
\end{equation}
Thus, to Gaussian order,
\begin{equation}
    \Omega(Q) \simeq e^{\Phi(\chi^*;Q)} \int \frac{d^r y}{(2\pi)^r} \exp\left[ -\frac{1}{2}y_AH^{AB}y_B \right] \left[ 1 + \cdots \right]
    = e^{\Phi(\chi^*;Q)} \frac{\mathcal C}{\sqrt{\det H}} \left[ 1 + \cdots \right].
    \label{eqA:TM_denominator_gaussian_integral}
\end{equation}
Here $\mathcal C$ denotes a source-independent normalization factor determined by the Gaussian integration measure and contour convention. Here and below, the ellipsis denotes non-Gaussian saddle corrections from cubic and higher derivatives of the exponent.

For the numerator, we define the shifted conserved quantity
\begin{equation}
    Q'^A = Q^A - n_k q_k^A - n_\ell q_\ell^A .
    \label{eqA:TM_Qprime_def}
\end{equation}
Then
\begin{equation}
    \Omega_{\neq k,\ell}(Q') = \int_{\Gamma}\frac{d^r\chi}{(2\pi i)^r} \exp\left[\Phi_{\neq k,\ell}(\chi;Q')\right],
    \label{eqA:TM_numerator_inverse}
\end{equation}
where
\begin{equation}
    \Phi_{\neq k,\ell}(\chi;Q') = \chi_{_A} Q'^A + \sum_{i\neq k,\ell}\log z_i(\chi).
    \label{eqA:TM_Phi_numerator}
\end{equation}
At the full-system saddle,
\begin{align}
    \Phi_{\neq k,\ell}(\chi^*;Q') &= \chi_{_A}^* \left( Q^A-n_k q_k^A-n_\ell q_\ell^A \right) + \sum_{i\neq k,\ell}\log z_i(\chi^*)
    \nn
    &= \Phi(\chi^*;Q) - \log z_k(\chi^*) - \log z_\ell(\chi^*) - n_k\chi_{_A}^*q_k^A - n_\ell\chi_{_A}^*q_\ell^A .
    \label{eqA:TM_reduced_Phi_at_saddle}
\end{align}
Unlike the denominator exponent, the reduced exponent is not stationary at $\chi^*$ because the occupations of modes $k$ and $\ell$ have been fixed. Its first derivative at $\chi^*$ is
\begin{align}
    \left. \frac{\partial\Phi_{\neq k,\ell}}{\partial\chi_{_A}} \right|_{\chi^*}
    &= Q'^A + \sum_{i\neq k,\ell} \left. \frac{\partial\log z_i}{\partial\chi_{_A}} \right|_{\chi^*}
    \nn
    &= Q^A - n_k q_k^A - n_\ell q_\ell^A - \sum_{i\neq k,\ell} q_i^A\langle n_i\rangle_*
    \nn
    &= Q^A - \sum_i q_i^A\langle n_i\rangle_* - (n_k-\langle n_k\rangle_*)q_k^A - (n_\ell-\langle n_\ell\rangle_*)q_\ell^A .
\end{align}
Using the saddle equation~\eqref{eqA:saddle_Q}, the first two terms cancel.
Therefore
\begin{equation}
    \left. \frac{\partial\Phi_{\neq k,\ell}}{\partial\chi_{_A}} \right|_{\chi^*} = - (n_k-\langle n_k\rangle_*)q_k^A - (n_\ell-\langle n_\ell\rangle_*)q_\ell^A .
    \label{eqA:TM_reduced_linear}
\end{equation}
It is useful to define
\begin{equation}
    R_{k\ell}^A = (n_k-\langle n_k\rangle_*)q_k^A + (n_\ell-\langle n_\ell\rangle_*)q_\ell^A .
    \label{eqA:Rkl_def}
\end{equation}
Then the numerator exponent expands as
\begin{equation}
    \Phi_{\neq k,\ell}(\chi;Q') = \Phi_{\neq k,\ell}(\chi^*;Q') - \delta\chi_{_A} R_{k\ell}^A + \frac{1}{2}\delta\chi_{_A} H^{AB}\delta\chi_{_B} +\cdots .
    \label{eqA:TM_numerator_Taylor}
\end{equation}
Using the same local contour parametrization $\delta\chi_{_A}=i y_A$, this becomes
\begin{equation}
    \Phi_{\neq k,\ell}(\chi;Q') = \Phi_{\neq k,\ell}(\chi^*;Q') - i y_A R_{k\ell}^A - \frac{1}{2}y_AH^{AB}y_B +\cdots .
    \label{eqA:TM_numerator_gaussian_y}
\end{equation}
Thus, to Gaussian order,
\begin{align}
    \Omega_{\neq k,\ell}(Q') &\simeq e^{\Phi_{\neq k,\ell}(\chi^*;Q')} \int \frac{d^r y}{(2\pi)^r} \exp\left[ -\frac{1}{2}y_AH^{AB}y_B - i y_A R_{k\ell}^A \right] \left[ 1 + \cdots \right]
    \nn
    &= e^{\Phi_{\neq k,\ell}(\chi^*;Q')} \frac{\mathcal C}{\sqrt{\det H}} \exp\left[ -\frac{1}{2} R_{k\ell}^A(H^{-1})_{AB}R_{k\ell}^B \right] \left[ 1 + \cdots \right].
    \label{eqA:TM_numerator_gaussian_integral}
\end{align}
In obtaining the second line, we used the standard Gaussian identity
\begin{equation}
    \int d^r u\, \exp\left[ -\frac{1}{2}u_AH^{AB}u_B - iJ^A u_A \right] = \frac{\mathcal C}{\sqrt{\det H}} \exp\left[ -\frac{1}{2}J^A(H^{-1})_{AB}J^B \right],
    \label{eqA:gaussian_identity}
\end{equation}
with $J^A=R_{k\ell}^A$. 

Dividing Eq.~\eqref{eqA:TM_numerator_gaussian_integral} by Eq.~\eqref{eqA:TM_denominator_gaussian_integral}, the leading Gaussian normalization cancels in the ratio. The residual determinant ratio from $H_{\neq k,\ell}$ is independent of $n_k,n_\ell$ at the order relevant for the mixed covariance and is absorbed into the normalization of the two mode distribution. We obtain
\begin{equation}
    \frac{\Omega_{\neq k,\ell}(Q')}{\Omega(Q)} \simeq \exp\left[ \Phi_{\neq k,\ell}(\chi^*;Q') - \Phi(\chi^*;Q) \right] \exp\left[ -\frac{1}{2} R_{k\ell}^A(H^{-1})_{AB}R_{k\ell}^B \right] \left[ 1 + \cdots \right].
\end{equation}
Using Eq.~\eqref{eqA:TM_reduced_Phi_at_saddle}, this becomes
\begin{equation}
    \frac{\Omega_{\neq k,\ell}(Q')}{\Omega(Q)} \simeq 
    \frac{e^{-n_k\chi_{_A}^*q_k^A}}{ z_k(\chi^*)} 
    \frac{e^{-n_\ell\chi_{_A}^*q_\ell^A}}{z_\ell(\chi^*)}
    \exp\left[ -\frac{1}{2} R_{k\ell}^A(H^{-1})_{AB}R_{k\ell}^B \right] \left[ 1 + \cdots \right].
    \label{eqA:TM_density_ratio_final}
\end{equation}
Substituting this into the exact two-mode law,
Eq.~\eqref{eqA:two_mode_exact}, gives
\begin{align}
    \mathbb{P}_{k\ell}(n_k,n_\ell\mid Q) &= \frac{1}{\mathcal{Z}_{k\ell}} \mathbb{P}_k^{(0)}(n_k\mid Q) \mathbb{P}_\ell^{(0)}(n_\ell\mid Q)
    \exp\left[ -\frac{1}{2} R_{k\ell}^A(H^{-1})_{AB}R_{k\ell}^B \right] \left[ 1 + \cdots \right].
    \label{eqA:TM_joint_gaussian}
\end{align}
where $\mathcal{Z}_{k\ell}$ normalizes the two mode distribution.

\subsection{Conservation-induced covariance projection}

Equation~\eqref{eqA:TM_joint_gaussian} is the conditional-Gaussian form of the two-mode law. Corrections beyond this form can modify local one-mode terms and higher cumulants. The leading connected correlation between two distinct modes is obtained from the mixed part of the quadratic compensation term.

Defining the occupation deviations
\begin{equation}
    \Delta_k = n_k-\langle n_k\rangle_*, \qquad
    \Delta_\ell = n_\ell-\langle n_\ell\rangle_* \,,
    \label{eqA:Delta_defs}
\end{equation}
Eq.~\eqref{eqA:Rkl_def} becomes
\begin{equation}
    R_{k\ell}^A = q_k^A\Delta_k+q_\ell^A\Delta_\ell .
    \label{eqA:Rkl_Delta}
\end{equation}
We also define the shorthand
\begin{equation}
    \Lambda_{ij} = q_i^A(H^{-1})_{AB}q_j^B \,.
    \label{eqA:Lambda_def}
\end{equation}
Then the quadratic form in Eq.~\eqref{eqA:TM_joint_gaussian} is
\begin{align}
    \frac{1}{2} R_{k\ell}^A(H^{-1})_{AB}R_{k\ell}^B &= \frac{1}{2}\Lambda_{kk}\Delta_k^2 + \frac{1}{2}\Lambda_{\ell\ell}\Delta_\ell^2 + \Lambda_{k\ell}\Delta_k\Delta_\ell .
    \label{eqA:compensation_expanded}
\end{align}
The first two terms depend only on a single occupation variable and can be absorbed into local one-mode corrections and the normalization $\mathcal{Z}_{k\ell}$. Edgeworth-type corrections from cubic and higher derivatives of the saddle exponent can also modify local terms and higher cumulants. They do not change the leading mixed Gaussian compensation term proportional to $\Lambda_{k\ell}\Delta_k\Delta_\ell$.

Expanding Eq.~\eqref{eqA:TM_joint_gaussian} to the order needed for the leading off-diagonal covariance, and retaining only the mixed term between
$\Delta_k$ and $\Delta_\ell$, gives
\begin{align}
    \mathbb{P}_{k\ell}(n_k,n_\ell\mid Q) &\simeq \mathbb{P}_k^{(0)}(n_k\mid Q) \mathbb{P}_\ell^{(0)}(n_\ell\mid Q) \left[ 1-\Lambda_{k\ell}\Delta_k\Delta_\ell+\cdots \right].
    \label{eqA:two_mode_mixed_expanded}
\end{align}
The omitted terms are either local in one of the two modes or beyond the conditional-Gaussian contribution relevant for the leading off-diagonal correlation.

We denote expectation with respect to the conditional two-mode law $\mathbb{P}_{k\ell}(n_k,n_\ell\mid Q)$ by $\langle\cdots\rangle_Q$. The conditional covariance between two distinct modes is
\begin{equation}
    \mathrm{Cov}_Q(n_k,n_\ell) = \langle n_k n_\ell\rangle_Q - \langle n_k\rangle_Q\langle n_\ell\rangle_Q,
    \qquad k\neq \ell .
    \label{eqA:cov_def}
\end{equation}
Since $\Delta_k$ and $\Delta_\ell$ differ from $n_k$ and $n_\ell$ only by constants, the same covariance can be written as
\begin{equation}
    \mathrm{Cov}_Q(n_k,n_\ell) = \langle \Delta_k\Delta_\ell\rangle_Q - \langle \Delta_k\rangle_Q \langle \Delta_\ell\rangle_Q .
    \label{eqA:cov_def_Delta}
\end{equation}

Let $\langle\cdots\rangle_0$ denote expectation with respect to the factorized leading law
\begin{equation}
    \mathbb P_k^{(0)}(n_k\mid Q)\,
    \mathbb P_\ell^{(0)}(n_\ell\mid Q).
\end{equation}
In particular,
\begin{equation}
    \langle \Delta_k\rangle_0=0,
    \qquad
    \langle \Delta_\ell\rangle_0=0,
    \qquad
    \langle \Delta_k^2\rangle_0=\sigma_{k,*}^2,
    \qquad
    \langle \Delta_\ell^2\rangle_0=\sigma_{\ell,*}^2 .
    \label{eqA:leading_Delta_moments}
\end{equation}
Using Eq.~\eqref{eqA:two_mode_mixed_expanded}, the one-mode shift of mode $k$ is
\begin{align}
    \langle \Delta_k\rangle_Q &= \sum_{n_k\in\mathcal{A}_k} \sum_{n_\ell\in\mathcal{A}_\ell} \Delta_k\, \mathbb P_k^{(0)}(n_k\mid Q) \mathbb P_\ell^{(0)}(n_\ell\mid Q) \left[ 1-\Lambda_{k\ell}\Delta_k\Delta_\ell+\cdots \right]
    \nn
    &= \langle \Delta_k\rangle_0 - \Lambda_{k\ell} \langle \Delta_k^2\rangle_0 \langle \Delta_\ell\rangle_0 +\cdots
    = 0+\cdots .
    \label{eqA:Delta_k_shift}
\end{align}
Similarly, $\langle \Delta_\ell\rangle_Q=0+\cdots$. Thus the product $\langle \Delta_k\rangle_Q\langle \Delta_\ell\rangle_Q$ does not contribute to the leading off-diagonal covariance in Eq.~\eqref{eqA:cov_def_Delta}.

The only term in Eq.~\eqref{eqA:two_mode_mixed_expanded} that couples the two modes is the mixed term proportional to $\Lambda_{k\ell}\Delta_k\Delta_\ell$. Its contribution to the remaining factor in Eq.~\eqref{eqA:cov_def_Delta} is
\begin{align}
    \langle \Delta_k\Delta_\ell\rangle_Q &= \sum_{n_k\in\mathcal{A}_k} \sum_{n_\ell\in\mathcal{A}_\ell} \Delta_k\Delta_\ell\, \mathbb{P}_k^{(0)}(n_k\mid Q) \mathbb{P}_\ell^{(0)}(n_\ell\mid Q) \left[ 1-\Lambda_{k\ell}\Delta_k\Delta_\ell+\cdots \right]
    \nn
    &= \langle \Delta_k\rangle_0 \langle \Delta_\ell\rangle_0 - \Lambda_{k\ell} \langle \Delta_k^2\rangle_0 \langle \Delta_\ell^2\rangle_0 +\cdots
    = - \Lambda_{k\ell}\sigma_{k,*}^2\sigma_{\ell,*}^2 +\cdots .
    \label{eqA:Delta_mixed_second_moment}
\end{align}
Combining Eqs.~\eqref{eqA:cov_def_Delta}, \eqref{eqA:Delta_k_shift}, and
\eqref{eqA:Delta_mixed_second_moment}, the leading off-diagonal covariance is
\begin{equation}
    \mathrm{Cov}_Q(n_k,n_\ell) = - \Lambda_{k\ell}\sigma_{k,*}^2\sigma_{\ell,*}^2 +\cdots 
    \qquad k\neq \ell .
    \label{eqA:offdiag_cov_Lambda}
\end{equation}
Using the definition $\Lambda_{k\ell} = q_k^A(H^{-1})_{AB}q_\ell^B$ we obtain
\begin{equation}
    \mathrm{Cov}_Q(n_k,n_\ell) = - \sigma_{k,*}^2\sigma_{\ell,*}^2 q_k^A(H^{-1})_{AB}q_\ell^B +\cdots ,
    \qquad k\neq \ell .
    \label{eqA:offdiag_cov_final}
\end{equation}
This is the leading conservation-induced intermode covariance. Including the intrinsic one-mode variance on the diagonal, the same result can be written in the compact conditional-Gaussian projection form
\begin{equation}
    C_{ij}^{Q} = \sigma_{i,*}^2\delta_{ij} - \sigma_{i,*}^2\sigma_{j,*}^2 q_i^A(H^{-1})_{AB}q_j^B .
    \label{eqA:projection_components}
\end{equation}
For $i\neq j$, Eq.~\eqref{eqA:projection_components} reduces to Eq.~\eqref{eqA:offdiag_cov_final}. For $i=j$, it gives the conditional-Gaussian projection of the local variance.

This form makes exact conservation manifest. Contracting with a conserved-charge vector gives
\begin{align}
    \sum_i q_i^A C_{ij}^{Q} &= \sigma_{j,*}^2 q_j^A - \sigma_{j,*}^2 \sum_i \sigma_{i,*}^2 q_i^A q_i^B (H^{-1})_{BC}q_j^C
    \nn
    &= \sigma_{j,*}^2 q_j^A - \sigma_{j,*}^2 H^{AB}(H^{-1})_{BC}q_j^C
    = 0 .
    \label{eqA:projection_null_left}
\end{align}
We used the definition~\eqref{eqA:Hessian_general}: $H^{AB} =\sum_i q_i^A q_i^B \sigma^2_{i,*}$ to go to the second equality. Similarly $\sum_j C_{ij}^{Q}q_j^A=0$. Thus, the projected covariance has no component along any exactly conserved direction.

\subsection{Linear observables and conservation-orthogonal projections}

The covariance projection can be applied directly to measured observables that are linear in the mode occupations. Recall that $n_i$ is the occupation number of mode $i$. We define two such observables by
\begin{equation}
    X=\sum_i f_i n_i,
    \qquad
    Y=\sum_i g_i n_i .
    \label{eqA:linear_observables}
\end{equation}
Here $f_i$ and $g_i$ are analysis coefficients assigned to mode $i$; they are not the microscopic counting weights $W_i(n_i)$. For example, in heavy-ion collisions,  $f_i=1$ gives a multiplicity-type observable in the selected set of modes, while $f_i=p_{T,i}$ gives a transverse-momentum-weighted observable.

Using Eq.~\eqref{eqA:projection_components}, the projected covariance of $X$ and $Y$ is
\begin{align}
    \mathrm{Cov}^{\rm cg}_Q(X,Y) &= \sum_{ij} f_i g_j C_{ij}^{Q}
    = \sum_i f_i g_i \sigma_{i,*}^2 - U_X^A(H^{-1})_{AB}U_Y^B ,
    \label{eqA:linear_cov_projection}
\end{align}
where
\begin{equation}
    U_X^A=\sum_i f_i\sigma_{i,*}^2 q_i^A,
    \qquad
    U_Y^A=\sum_i g_i\sigma_{i,*}^2 q_i^A .
    \label{eqA:U_defs}
\end{equation}
The vectors $U_X^A$ and $U_Y^A$ measure how strongly the observables overlap with the exactly conserved quantities. If $X$ and $Y$ are built from disjoint sets of modes, the diagonal term in Eq.~\eqref{eqA:linear_cov_projection} is absent. In that case, the long-range conservation-induced covariance is
\begin{equation}
    \mathrm{Cov}^{\rm cons}_Q(X,Y) = - U_X^A(H^{-1})_{AB}U_Y^B .
    \label{eqA:linear_cons_cov}
\end{equation}

This form gives a direct prescription for constructing observables with no leading overlap with the conserved quantities. We seek modified coefficients $f_i^\perp$ satisfying
\begin{equation}
    \sum_i \sigma_{i,*}^2 f_i^\perp q_i^C =0
    \qquad \text{for all } C .
    \label{eqA:orthogonality_condition}
\end{equation}
Starting from a given set of analysis coefficients $f_i$, subtract a linear  combination of conserved-charge vectors,
\begin{equation}
    f_i^\perp = f_i-q_i^A \lambda_A .
    \label{eqA:orthogonal_ansatz}
\end{equation}
Imposing Eq.~\eqref{eqA:orthogonality_condition} gives
\begin{align}
    0 &= \sum_i \sigma_{i,*}^2 f_i q_i^C - \lambda_A \sum_i \sigma_{i,*}^2 q_i^A q_i^C
    = U_X^C-H^{CA}\lambda_A .
\end{align}
Hence
\begin{equation}
    \lambda_A=(H^{-1})_{AB}U_X^B ,
\end{equation}
and therefore
\begin{equation}
    f_i^\perp = f_i-q_i^A(H^{-1})_{AB}U_X^B .
    \label{eqA:orthogonal_coeff}
\end{equation}
Thus the projected observable is
\begin{equation}
    X^\perp=\sum_i f_i^\perp n_i \,.
\end{equation} 
Using Eqs.~\eqref{eqA:U_defs}, its conserved-charge overlap vanishes,
\begin{align}
    U_{X^\perp}^C =  \sum_i \sigma_{i,*}^2 f_i^\perp q_i^C &= \sum_i \sigma_{i,*}^2 f_i q_i^C - \sum_i \sigma_{i,*}^2 q_i^Cq_i^A (H^{-1})_{AB}U_X^B
    \nn
    &= U_X^C - H^{CA}(H^{-1})_{AB}U_X^B = 0 .
    \label{eqA:orthogonality_check}
\end{align}
Therefore, for any linear observable $Y$,
\begin{align}
    \mathrm{Cov}^{\rm cg}_Q(X^\perp,Y) &= \sum_i f_i^\perp g_i\sigma_{i,*}^2 - U_{X^\perp}^A(H^{-1})_{AB}U_Y^B
    = \sum_i f_i^\perp g_i\sigma_{i,*}^2 .
    \label{eqA:projected_covariance}
\end{align}
Therefore, the finite-rank conservation contribution vanishes once one observable is projected. If the two observables are built from disjoint sets of modes, the diagonal term is absent, and hence
\begin{equation}
    \mathrm{Cov}^{\rm cons}_Q(X^\perp,Y)=0 \,.
\end{equation}
Thus the projected observable has no leading conservation-induced long-range covariance with any disjoint linear observable. Local statistical correlations and dynamical correlations, if present, are not removed by this construction. This provides a practical way to design weighted observables that are orthogonal to exact conservation laws before comparing with dynamical correlations.

\end{document}